# All-carbon multi-electrode array for real-time in vitro measurements of oxidizable neurotransmitters

*Federico Picollo[1,2,3]\*, Alfio Battiato[2,1,3], Ettore Bernardi[2,1,3], Marilena Plaitano[2], Claudio Franchino[4,3], Sara Gosso[4,3], Alberto Pasquarelli[5], Emilio Carbone[4,3], Paolo Olivero[2,1,3] and Valentina Carabelli[4,3]*

\*Corresponding Author: E-mail: picollo@infn.it

[1] Istituto Nazionale di Fisica Nucleare (INFN) Sez. Torino; via P. Giuria 1, 10125, Torino, Italy

[2] Physics Department and "NIS" Inter-departmental Centre - University of Torino; via P. Giuria 1, 10125, Torino, Italy

[3] Consorzio Nazionale Inter-universitario per le Scienze fisiche della Materia (CNISM) Sez. Torino, Italy

[4] Drug Science and Technology Department and "NIS" Inter-departmental Centre - University of Torino; Corso Raffaello 30, 10125, Torino, Italy

[5] Institute of Electron Devices and Circuits - University of Ulm – Ulm; Albert Einstein Allee 45, 89069, Germany

**Abstract**

We report on the ion beam fabrication of all-carbon multi electrode arrays (MEAs) based on 16 graphitic micro-channels embedded in single-crystal diamond (SCD) substrates. The fabricated SCD-MEAs are systematically employed for the *in vitro* simultaneous amperometric detection of the secretory activity from populations of chromaffin cells, demonstrating a new sensing approach with respect to standard techniques.

The biochemical stability and biocompatibility of the SCD-based device combined with the parallel recording of multi-electrodes array allow: i) a significant time saving in data collection during drug screening and/or pharmacological tests over a large number of cells, ii) the possibility of comparing altered cell functionality among cell populations, and iii) the repetition of acquisition runs over many cycles with a fully non-toxic and chemically robust bio-sensitive substrate.

Synaptic transmission is regulated by vesicle exocytosis. This process occurs when a neuron directs the content of secretory vesicles from the pre-synaptic terminal into the inter-synaptic space, thus delivering neurotransmitter molecules that activate post-synaptic receptors of neighbouring neurons. This release mechanism is common to all neurons and is shared by neuroendocrine cells [1,2], including the chromaffin cells of the adrenal medulla. Such cells release catecholamines (e.g. adrenaline and noradrenaline) following splanchnic nerve stimulation [3]. As such, chromaffin cells are widely used as a model system to study the molecular events underlying exocytosis. The release of oxidazable molecules from individual excitable cells is commonly measured using amperometry, an electrochemical technique that allows resolving the kinetics of single secretory events with a high time resolution [4-7].

Since few decades carbon fibre microelectrodes (CFEs) [4,8] are the most widely employed electrodes to detect the quantal release of catecholamines and still represent the reference technique in this field. Though, parallel recordings of catecholamine release from populations of chromaffin cells still remains a crucial issue that cannot be addressed by means of conventional CFE-based setups. Recently, in order to overcome this limitation, planar multi-electrode devices fabricated from either indium tin oxide (ITO) [9-11], diamond-like carbon (DLC) [12], boron-doped nanocrystalline diamond [13], noble metal (Au, Pt) [14,15] and CMOS silicon-based chips [16] have been successively developed for the simultaneous recording of catecholamine release from chromaffin cell populations.

In the present work, we describe the parallel fabrication of single-crystal-diamond substrates for the realization of micro-electrode array sensors (SCD-MEAs) based on graphitic microchannels, and demonstrate their capabilities in sensing *in vitro* the secretory activity of cultured chromaffin cells. These devices allow to overcome several critical issues of standard devices such as: mechanical and chemical stability over long measurement periods, substrate transparency, biocompatibility, parallel multi-electrode recordings and the possibility of directly culturing the living cells on the substrate.

**Results**

**Device microfabrication.** The SCD-MEAs were realized using optical-grade single-crystal artificial diamond substrates by means of an advanced MeV ion beam lithography technique, whose effectiveness was previously demonstrated by the proof-of-concept single-electrode device reported in [17].

Two samples were implanted at room temperature with a 1.2 MeV $He^+$ ion broad beam (see the "Methods" section for further details). The implantation of MeV ions induces structural damage in matter mainly at end of the penetration range of the ions, where the cross section for nuclear collisions is strongly enhanced. The strongly non-uniform depth profile of the damage density can be evaluated in first approximation as the product between the linear damage density (as evaluated with the SRIM Monte Carlo code [18], see Fig. 1a and the "Methods" section for further details) and the implantation fluence. Such a linear approximation, although disregarding cumulative damage effects such as self-annealing and defect-defect interactions, provides a sufficiently accurate prediction of the depth and thickness of the sub-superficial layer where the high damage density overcomes a critical threshold, thus inducing a full amorphization of the material [19-21]. The implantation fluence ($F = 1.5 \times 10^{17}$ $cm^{-2}$) was chosen to determine the formation of a ~250 nm thick and ~2 μm deep fully amorphized layer below the sample surface.

The three-dimensional geometry of the implanted structures was defined by the combination of two masking systems, as shown in Figure 1. The first system consists of an array of variable-thickness copper masks directly deposited over the sample surface which modulate the penetration depth of the ions, thus ensuring the connection of the sub-superficial amorphized structures with the sample surface at their endpoints [22-24]. 32 square (20×20 $μm^2$ and 3 μm maximum thickness) metal depositions were defined to this scope. The second system consists of a free-standing mask realized in a 15 μm thick aluminium foil microfabricated with high-power laser ablation. This mask is thick enough to fully stop the broad MeV ion beam, thus allowing the definition of the lateral geometry of the amorphized regions. The two masking systems were suitably aligned to simultaneously define in each sample 16 amorphous-carbon micro-channels with emerging end-points.

After the implantation, a high temperature thermal annealing (950 °C for 2 hours in high vacuum) was performed to induce the permanent conversion of the amorphized regions to a graphite-like phase. As a consequence, 16 graphitic micro-channels (width: ~20 µm, length: 1.4-1.9 mm, thickness: ~250 nm) were obtained at a depth of ~2 µm in each substrate, with surface-emerging endpoints, as schematically shown for a single channel in Figures 1c and 1d. As shown in Figures 2b and 2c, each device consists of 16 sub-superficial graphitic micro-channels converging into a central region where their endpoints (i.e. the sensing electrodes) are exposed to the sample surface. As a consequence of the geometrical elongation of its thickness as it emerges in contact with surface [22], each ~250 nm thick channel terminates with a ~20×3.5 µm$^2$ sensing electrode at the sample surface, and the 16 electrodes are arranged on a 4×4 square grid with 200 µm spacing. At the substrate periphery, the other emerging endpoints of the micro-channels provide electrical contacts for bonding to a dedicated chip carrier. The adopted geometry of the electrodes arrays allowed to rule out the "cross talking" between signals detected at different electrodes since the detection of the out-diffusion of molecules from the cell is impossible at these cell-electrode distances. Indeed, no indications of simultaneous detection of the same releases events from different electrodes were ever found during measurements.

The chip carrier was specifically designed to perform *in vitro* measurements on cells directly cultured on the microfabricated diamond substrate and was therefore equipped with a 4 ml perfusion chamber. The chip carrier was directly plugged into the front-end electronics and controlled by an acquisition software developed in LabView$^{TM}$ environment [13,17] (see the "Methods" section for further details). Biocompatible epoxy resin was employed to seal the chip in a long-term inert and stable packaging. The chip assembly was mounted on an inverted transmission microscope, thus allowing the monitoring of the cell condition and position during the amperometric recording in virtue of the overall transparency of the diamond substrate (see Figs. 2a and 2b).

**Electrical characterization.** The graphitic microchannels of both devices were electrically characterized before performing the cell sensing. The linearity of the I-V curves (similar to what reported in [22-24], not reported here) indicates an ohmic conduction within the sub-superficial electrodes with resistance values ranging from 5 k$\Omega$ to 9 k$\Omega$, depending upon geometrical factors. If such factors are properly accounted for, these resistance values correspond to resistivity values of $\rho \sim$ 1-2 m$\Omega$ cm, comparable with those of standard polycrystalline graphite ($\rho \sim$ 1.3 m$\Omega$ cm) [25]. Due to fabrication issues, both devices presented 1 or 2 channels with significantly higher resistance values (i.e. >10 M$\Omega$). These channels were discarded from subsequent data analysis.

The electrochemical sensing performance of the SCD-MEAs was tested by cyclic voltammetry in liquid environment employing either a Tyrode physiological solution containing (in mM) 128 NaCl, 2 MgCl$_2$, 10 glucose, 10 HEPES, 10 CaCl$_2$ and 4 KCl, or a salt-free solution containing only adrenaline (100 mM). A triangular voltage waveform with a scan rate of 20 mV s$^{-1}$ ranging from -0.5 V to +1.1 V was applied to the graphitic micro-electrodes with respect to an Ag/AgCl quasi-reference electrode immersed in the bath solution. No redox activity was observed using the Tyrode solution within the anodic range of the hydrolysis window, i.e. up to bias voltages of +900 mV. Under these conditions, a leakage current lower than 10 pA was measured at +600 mV bias voltage. The width of the electrochemical window of the graphitic microelectrodes allows the detection of the oxidation peak of adrenaline, located between +500 mV and +800 mV, as shown in Figure 3.

On the basis of these results, the optimal bias voltage for the subsequent amperometric measurement was set to +800 mV, i.e. in correspondence of the maximum value of the ratio between the voltammetric signals of adrenaline oxidation and water hydrolysis.

**Amperometric recordings of exocytotic events.** Bovine chromaffin cells were plated over the SCD-MEAs following the procedures reported in the "Methods" section. The direct incubation of the cells over the devices allowed achieving the best performances of the sensors while guaranteeing simultaneous amperometric acquisitions.

The exocytosis from populations of chromaffin cells was chemically stimulated by applying a 30 mM KCl-enriched Tyrode solution, which depolarizes the cells and causes massive vesicle fusion and secretion. All sensing electrodes were simultaneously polarized at +800 mV, thus drastically accelerating data collection with respect to conventional single-electrode techniques (i.e. CFE) and allowing non-invasive repeated measurements over long acquisition times. Typical sequences of amperometric spikes evoked by the KCl solution and collected by the 16 microelectrodes over different acquisitions are shown in Figure 4. The chrono-amperometric recordings were collected for 120 s after cell stimulation with a sampling rate 4 kHz per each channel. As expected, the peak-to-peak noise level of the acquired signals was 10 pA. The secretory activity was recorded in most of the microelectrodes, while some electrodes were silent due to the inhomogeneous cell coverage of the electrode surface (see Fig. 2c). In Figure 4 (bottom-right) we report subsequent time-zooms of the chrono-amperogram collected from channel #9 of device #2, as well as a single amperometric spike. In the single spike the typical kinetic features of exocytotic process are visible with excellent signal sensitivity and time resolution, such as the opening and expansion of the fusion pore indicated by the "foot" signal preceding the main peak [26]. Since the time shape and intensity of each spike carries information on the sequential dynamics of neurotransmitter release, the SCD-MEAs allow a thorough investigation of the release mechanisms from the cell population under study. With the aim of quantifying the kinetic features of the amperometric signals, we measured the following spike parameters: maximum oxidation current $I_{max}$, total electrical charge $Q$, half-width time $t_m$, slope $m$ and rise time $t_p$ of the rise front.

**Exocytic spikes analysis.** Multiple simultaneous recordings from 70 cells (corresponding to over 3,000 spikes) were processed by means of a dedicated software [27], thus allowing the implementation of a statistical approach to data analysis. Signals affected by diffusion process and associated with release events occurring in cells far from the electrodes (corresponding to values of $m < 1$ nA s$^{-1}$ and/or current intensities lower than 15 pA), as well as multiple events, were discarded. Generally, on the basis of these selection criteria more than 10 spikes could be isolated from each recording (240 s of mean duration), while recordings with less than 5 spikes were not considered due to lack of statistical significance.

In the analysis of spike parameters, our first goal was to verify that the amperometric signals could be detected with comparable sensitivity by the different electrodes of the 2 SCD-MEAs employed in this study. Figure 5a shows the distribution of the charge values $Q$ recorded with the different electrodes of the 2 devices. Each point corresponds to the $Q$ average value obtained by a single recording of consecutive amperometric spikes from a single cell from the indicated electrode, and it is associated to an uncertainty corresponding to the standard deviation of the relevant data distribution. It is worth remarking that analogous processing can be implemented for the other above-mentioned parameters, thus providing significant insight in the observed release mechanisms. The variability of the measured $Q$ values in all the electrodes can be attributed to the variability of the cell population, as demonstrated by the fact that different acquisitions recorded from the same electrode display a dispersion deriving from the intrinsic variability of the cells under analysis. As expected, such an intrinsic dispersion is statistically compatible with the inter-electrode dispersion. This was demonstrated by means of ANOVA analysis followed by Bonferroni *post hoc* comparison, thus confirming that each electrode from both devices detects exocytotic events with statistically compatible sensitivity and time resolution. All the characteristic spike parameters ($I_{max}$, $Q$, $t_m$ and $t_p$) were considered in the analysis and no statistically significant variations among the electrodes were found at a 5% confidence level, and a 1% confidence level for the $m$ parameter. Therefore, in the following analysis, the data recorded from different electrodes of the 2 devices will be gathered in a single dataset.

With the purpose of assessing the performance of the SCD-MEAs, two further experiments were carried out: 1) a comparison between the sensitivities of the SCD-MEAs and of CFEs used in the same cell plating conditions and over the same cell population, and 2) a comparison of the chrono-amperograms obtained from the cell population (see Fig. 4) with the recordings from a single chromaffin cell. In the former tests, CFEs were employed to detect exocytic events from cells incubated on fully equivalent diamond substrates for sake of consistency, while in the latter tests individual cells were suitably positioned on single microelectrodes of the SCD-MEAs by means of a patch-clamp micromanipulator.

Table 1 summarizes the characteristic parameters of the quantal exocytotic events obtained from the above mentioned experiments, as well as reference literature values. Given the proved homogeneity of the electrodes performance, these parameters were obtained by averaging of the mean values of the spike parameters obtained from each cell and estimating the corresponding uncertainty with the standard error associated to the calculated averages. The statistical significance of all the measured parameters with the corresponding reference values (CFE and single-cell recordings, literature values) was estimated by the Z-test [28] at a 5% confidence level. No statistically significant differences were observed between these measurements, thus confirming a close correspondence of the SCD-MEA recordings with state-of-the-art methods, while offering the significant advantages of multiple simultaneous recording over long times in a robust and bio-chemically stable environment.

The analysis of the $Q$ parameter was extended by determining the distribution of the cubic root charge values $Q^{1/3}$ pooled from all electrodes of both devices (see Fig. 5b). The distribution of the $Q^{1/3}$ values is bimodal, with two Gaussian peaks centred respectively at $(1.01 \pm 0.05)$ pC$^{1/3}$ and $(1.34 \pm 0.09)$ pC$^{1/3}$, suggesting the presence of two populations of granules with different size. The effect of the variable cell-electrode distance on the measured signal distributions (i.e. the diffusional loss of released cargo molecules) could be ruled out by excluding from the data analysis spikes characterized by slope values of $m < 1$ nA s$^{-1}$ and/or maximum current intensities of $I_{max} < 15$ pA. Therefore, the observation of a bimodal distribution can be reasonably ascribed to the physiological properties of the probed cells. The distribution of the vesicles diameter was derived from the $Q^{1/3}$ values using Faraday law [31] and assuming spherical shape and uniform concentration of adrenaline inside the granules:

$$d = 2\left(\frac{3Q}{4\pi nFM}\right)^{1/3} \qquad (1)$$

where $Q$ is the total charge for vesicle, $n$ is the number of electrons involved in the electrochemical oxidation ($n = 2$ for catecholamines [31]), $F$ is the Faraday constant and $M$ is the adrenaline concentration ($M = 0.75$ M [29]). From equation (1) two vesicle size distributions peaked at diameters of ∼240 nm and ∼320 nm were obtained, in good agreement with similar values reported for mouse chromaffin cells derived from electron microscopy studies [32].

**Discussion**

We demonstrated that the newly developed SCD-MEAs are highly sensitive and perfectly suitable for the parallel amperometric recordings from populations of living chromaffin cells. The SCD-MEA guarantees high levels of reliability of electrode performances and marked stability with time (6-12 months of stable functioning). The SCD-MEA can resolve amperometric signals with state-of-the-art sensitivity and high time resolution [33], while offering the significant advantage of allowing parallel synchronous acquisitions from a large number of cultured cells. In addition, the parallel production technique offers promising perspectives in the systematic production of large number of devices using standard ion implanters. Thus, the proposed strategy will allow to easily increase the number of sensing microelectrodes, anticipating attractive applications directed to the analysis of complex neuronal networks during synapses maturation, the functioning of intact neuroendocrine tissues and drug screening tests on various cell types in combination with optical fluorescence measurements. The newly developed fabrication process allows a further miniaturization of the device, which will open the possibility of studying quantal secretory events with sub-cellular spatial resolution (1-3 μm$^2$) [34].

**Methods**

*Ion beam implantation*: the diamond-based MEAs were realized in high-purity monocrystalline CVD (Chemical Vapour Deposition) diamond substrates produced by ElementSix$^{TM}$ (UK). The substrates are 4.5×4.5×0.5 mm$^3$ in size, cut along the (100) crystalline direction and optically polished on the two opposite large faces. The crystals are classified as type IIa, with substitutional nitrogen and boron concentrations below 1 ppm and 0.05 ppm, respectively. Such low impurity concentrations ensure a good transparency of the substrates in the visible spectrum, up to the near UV.

1.2 MeV He$^+$ ion beam implantation was carried out at the AN2000 accelerator facility of the INFN Legnaro National Laboratories. A fluence of 1.5×10$^{17}$ cm$^{-2}$ was uniformly delivered over a beam spot of 6×6 mm$^2$. The ion current was ~400 nA.

The damage profile in the diamond lattice was simulated by SRIM-2013.00 code version [18], setting the same parameters (displacement energy, density, calculation mode, etc.) as mentioned in previous works [22] (see Fig. 1). The high damage density induced by ion implantation promotes the conversion of the diamond lattice into an amorphous phase within a layer which is located ~2 μm below the sample surface. The employed fluence was chosen to overcome a critical damage density (usually referred to as "graphitization threshold") in correspondence of the Bragg peak. Such threshold has been estimated with an effective value of $9\times10^{22}$ vacancies cm$^{-3}$ in the above-mentioned linear approximation.

*Acquisition electronic chain*: the front-end electronics consists of 16 low noise trans-impedance amplifiers having an input bias current of ~2 pA and a gain of 100 MΩ, followed by Bessel low-pass filters of the 6$^{th}$ order with a cut-off frequency of 1 kHz. The filtered signals are then acquired by a 16-bit analog/digital converter working over an input range of ±10 V at a sampling rate of 4 kHz per channel. A buffered 16-bit digital/analog converter (DAC) provides the bias voltage for the catecholamine oxidation, over the common quasi-reference Ag/AgCl electrode immersed into the working solution. Both the data acquisition and oxidation potential are controlled by software developed in LabView$^{TM}$ environment.

*Cell culture:* bovine chromaffin cells were isolated as described in [34]. Adrenal glands from 6- to 18-month-old cows were rapidly transported to the culture unit in Locke buffer containing (mM): 154 NaCl, 5.6 KCl, 3.5 NaHCO$_3$, 5.6 glucose and 10 Hepes (pH 7.3 with NaOH). Medulla digestion was performed by injecting the following solution for 3 times at 30 min intervals: 3 ml of Locke solution mixed with 0.2% collagenase (Sigma), 1.7% hyaluronidase (Sigma), 0.5% BSA (Sigma) and 0.15% DNase I (Sigma). After separation of the medulla from the cortex, the tissue suspension was filtered with a nylon mesh (217 μm pore) and centrifuged at a radial acceleration of 71.5 *g* for 12 min at room temperature. The supernatant and the upper pellet of erythrocytes were removed, while the pellet of chromaffins was re-suspended in DMEM (GIBCO, Grand Island, NY, USA) and filtered with a nylon mesh (80 μm pore). Cells were plated at a high surface density of ~$1.5\times10^4$ cells $mm^{-2}$ and incubated at 37 °C in a water-saturated 5% $CO_2$ atmosphere. The culture medium contained: DMEM, fetal calf serum (15%, GIBCO), penicillin (50 IU $ml^{-1}$) and streptomycin (50 μg $ml^{-1}$; GIBCO). Cells were used within 2–4 days after plating. All experiments were conducted in accordance with the guidelines on Animal Care established by the Italian Minister of Health and were approved by the local Animal Care Committee of Turin University.

**Acknowledgements**

The authors thank Ottavio Giuliano for his technical support. This work is supported by the following projects: "DiNaMo" (young researcher grant, project n° 157660) by National Institute of Nuclear Physics; FIRB "Futuro in Ricerca 2010" (CUP code: D11J11000450001) funded by MIUR, "A.Di.N-Tech." (CUP code: D15E13000130003) funded by the University of Torino and "Compagnia di San Paolo". The MeV ion beam lithography activity was performed within the "Dia.Fab." experiment of the INFN Legnaro National Laboratories.

**Display items – Figure captions**

Figure 1. **Deep ion beam lithography of synthetic diamond.** a) Vacancy density profile induced by 1.2 MeV $He^+$ ions implanted in diamond at a fluence of $1.5\times10^{17}$ $cm^{-2}$. The horizontal line indicates the graphitization threshold, while the patterned rectangle highlights the thickness of the graphitic layer formed upon thermal annealing. Schematics (not to scale) of the fabrication a sub-superficial graphitic channel: three-dimensional view of the masked diamond (b), zoom of three-dimensional (c) and cross-sectional (d) view of a single channel with emerging end-points. The lateral features of the electrode are defined by the aperture in the non-contact mask (in gray), while its depth profile is defined by the contact variable-thickness metallic mask (in yellow).

Figure 2. **SCD-MEA layout**. a) Picture of the sensor mounted on the acquisition electronic board. b) Optical micrograph of the diamond sample soldered on the chip carrier. c) Optical transmission micrograph of living chromaffin cells cultured on top of a SCD-MEA. Chromaffin cells cover most of the available microelectrodes. The emerging endpoints of the graphitic channels (i.e. the surface electrodes) are circled in red and sub-superficial insulated graphitic path are highlighted with blue dashed lines. d) Optical transmission micrograph of a living chromaffin cell in close proximity to the active region of a microelectrode. The emerging endpoint of the graphitic channel is highlighted with red solid lines while the sub-superficial insulated graphitic path is highlighted with blue dashed lines.

Figure 3. **Cyclic voltammetry characterization.** Cyclic voltammetric scans at 20 mV s$^{-1}$ rate of a -0.5 ÷ 1.1 V ramp voltage applied to all microelectrodes with respect to the quasi-reference Ag/AgCl electrode in the presence of Tyrode buffer (solid black lines) and 100 mM adrenaline solution (solid red lines). All electrodes show the characteristic features of water oxidation at a voltage higher then +0.850 V, while detecting the oxidation of adrenaline at voltages between +500 mV and +800 mV, with the exception of electrode #2.

Figure 4. **Exocytic events recordings from chromaffin cells.** Typical chrono-amperogrametric recordings from individual cultured chromaffin cells positioned in close proximity to the respective graphitic electrodes after stimulation with a KCl-enriched solution. Amperograms were collected simultaneously by the 16 electrodes polarized at +800 mV bias with respect to the quasi-reference electrode. In the bottom right, details of the signals detected by electrode #9 are shown at increasing time scale magnification. In the green

rectangle, a single amperometric spike is clearly anticipated by a "foot" current associated with the opening and expansion of the fusion pore.

Figure 5. *Q* **parameter analysis.** a) Recorded charge *Q* over the different electrodes of the 2 SCD-MEAs; each datapoint represents the average value of the charge measured from the spikes recorded from an electrode during a measurement run, while the uncertainty bar is given by the standard error of the relevant dataset. b) cubic root charge histogram distribution fitted by a double Gaussian function: the peaks position are $(1.01 \pm 0.05)$ pC$^{1/3}$ and $(1.34 \pm 0.09)$ pC$^{1/3}$, respectively.

|  | $I_{max}$ (pA) | $Q$ (pC) | $Q^{1/3}$ (pC$^{1/3}$) | $t_m$ (ms) | $m$ (nA s$^{-1}$) | $t_p$ (ms) | # spikes | # cell |
|---|---|---|---|---|---|---|---|---|
| SCD-MEA: cultured cells | 74 ±5 | 1.56 ±0.09 | 1.06 ±0.02 | 17.0 ±0.7 | 21 ±2 | 8.3 ±0.4 | 3003 | 70 |
| CFE (cultured cells on diamond plate) | 82 ±13 | 1.4 ±0.2 | 0.97 ±0.05 | 17.2 ±1.6 | 33 ±6 | 8.0 ±0.8 | 495 | 14 |
| SCD-MEA: single cell | 99 ±23 | 1.5 ±0.2 | 1.10 ±0.04 | 14.3 ±1.5 | 40 ±13 | 6.0 ±0.7 | 307 | 10 |
| Literature[29] and §[30] | 73 ±3 | 1.40 ±0.06 | 1.06 § ±0.04 | 16.0 ±0.5 | 24.2 ±1.1 | 18.2 ±1.0 | 778 | 12 |

Table 1. Mean values and relevant uncertainties of the main parameters describing the amperometric spikes recorded with the SCD-MEA, as compared with corresponding values obtained by: i) CFEs on chromaffin cells cultured on a diamond plate, ii) single cells positioned by patch-clamp pipette on the electrodes the SCD-MEA and iii) literature [29, 30]

**Author contribution statement**

F.P, A.B. and E.B prepared the sample. S.G, M.P. and V.C. did the experiments. A.P. realized the acquisition hardware and software. C.F. prepared the cell culture. P.O. and E.C. designed the experiment. F.P and A.B. wrote the manuscript with input from all coauthors. All authors reviewed the manuscript.

**Additional Information**

The authors declare no competing financial interests.

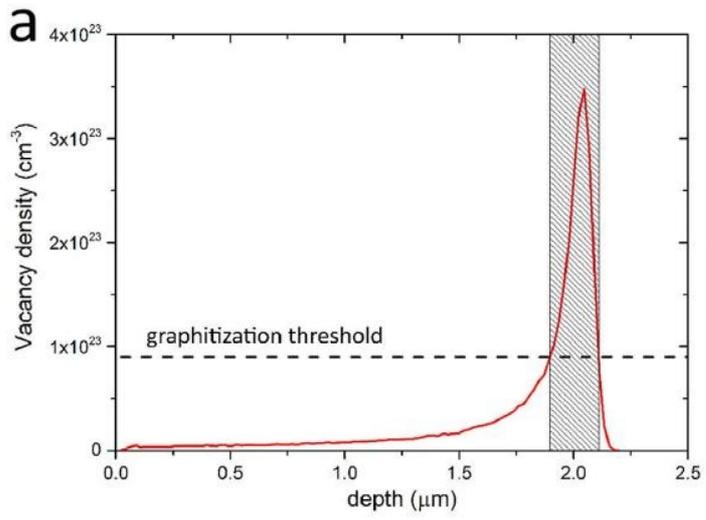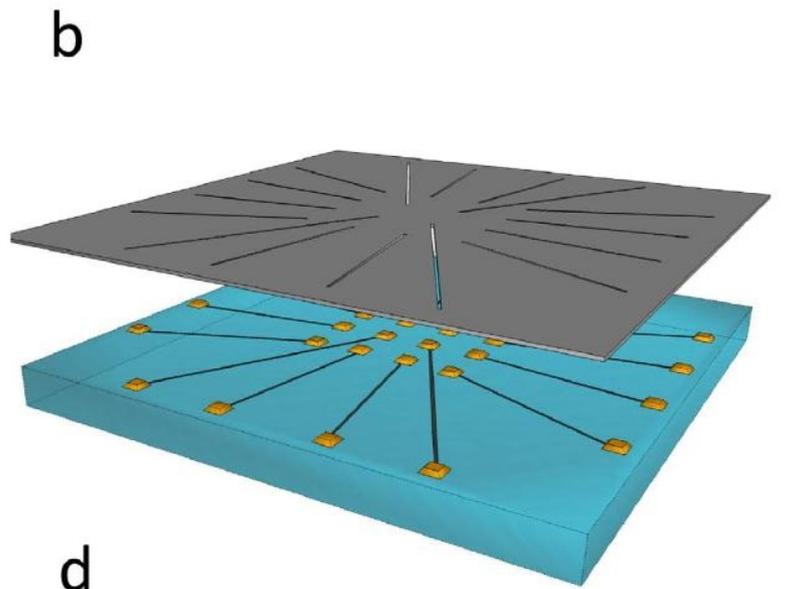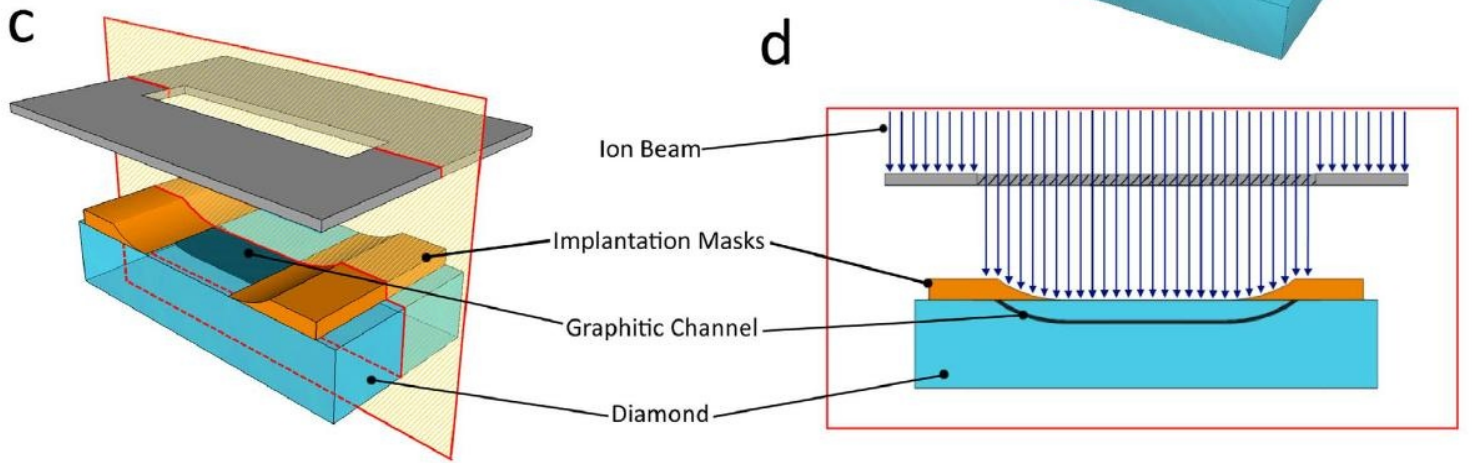

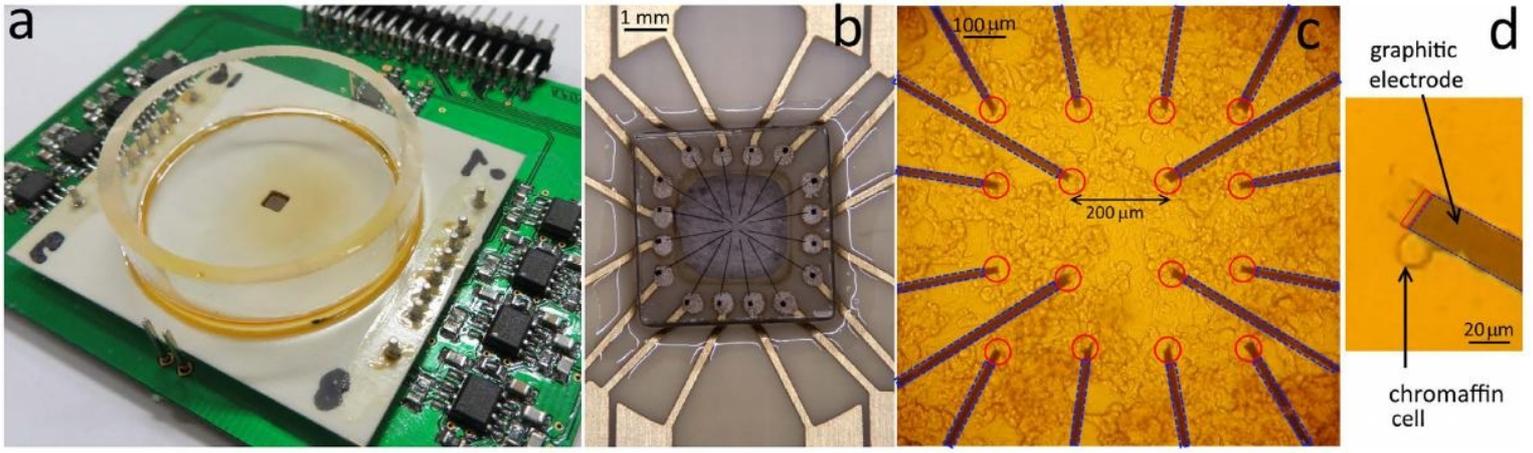

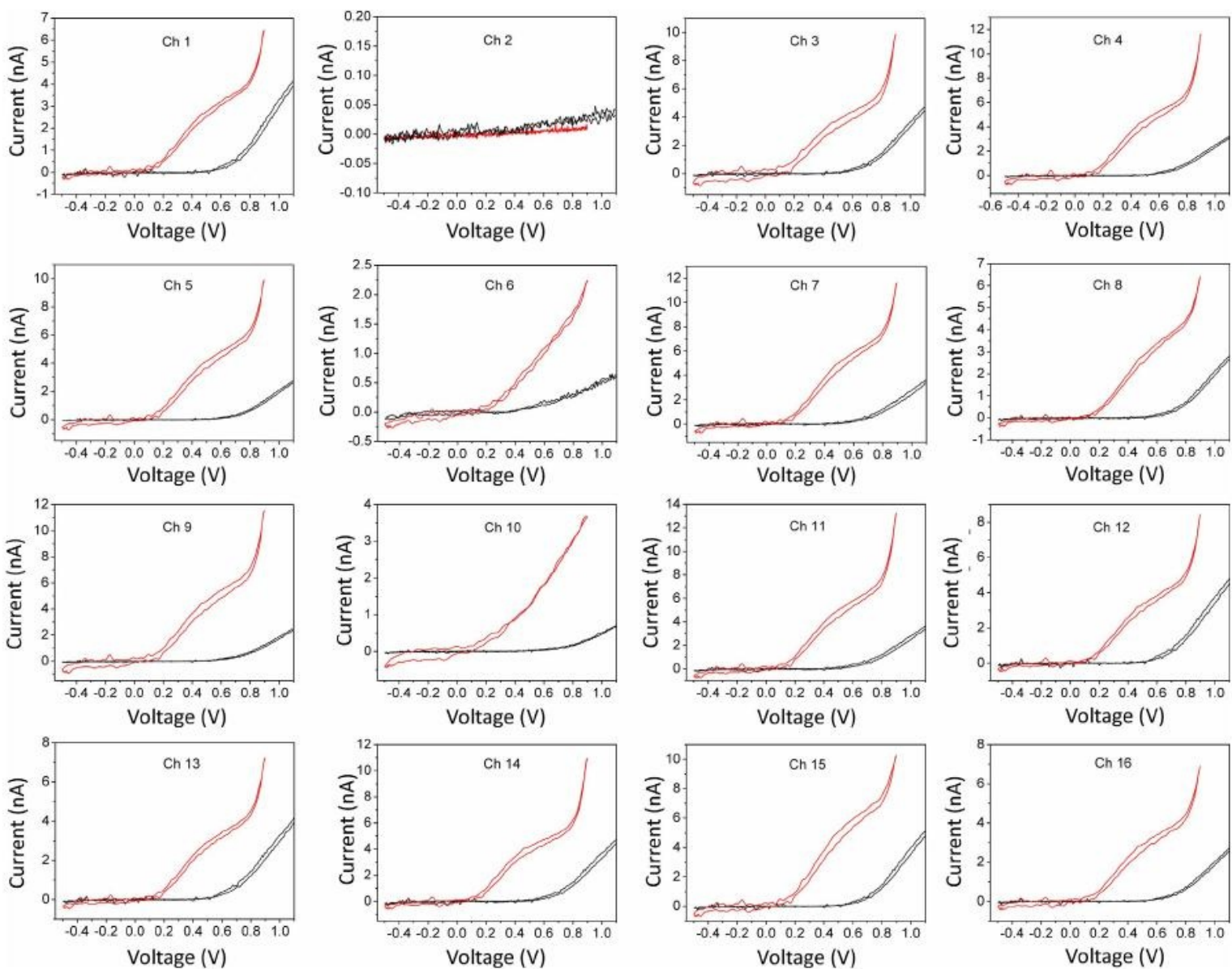

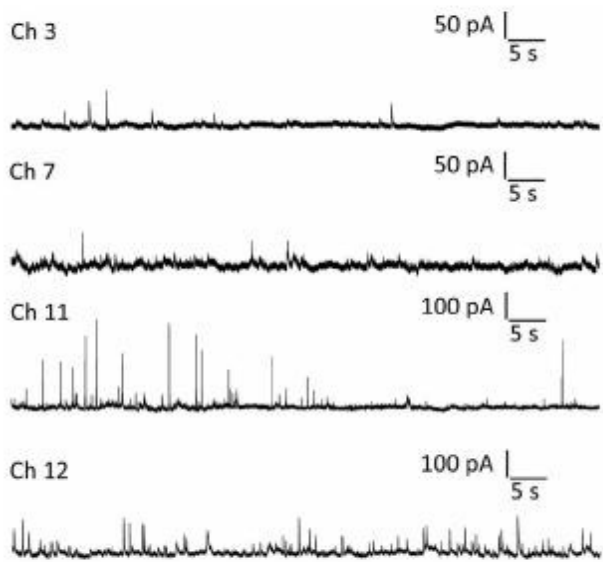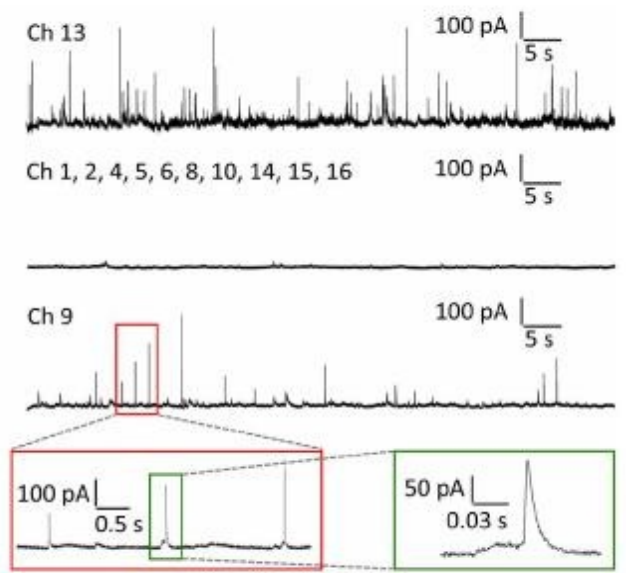

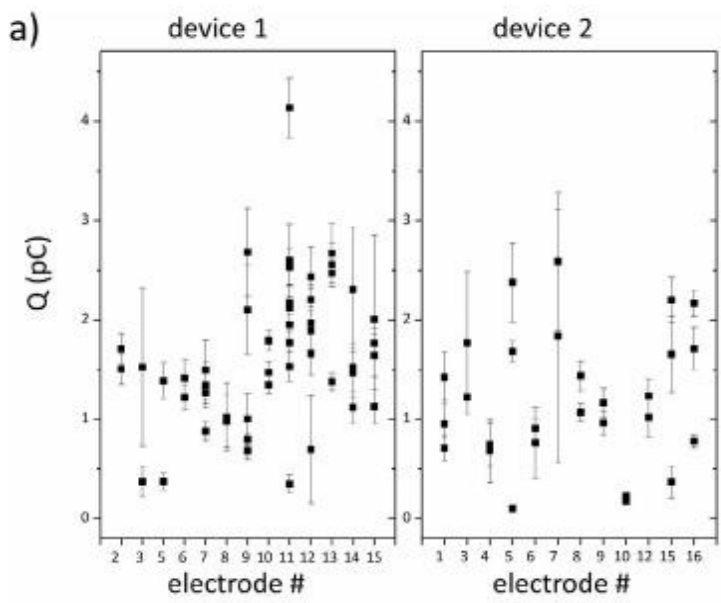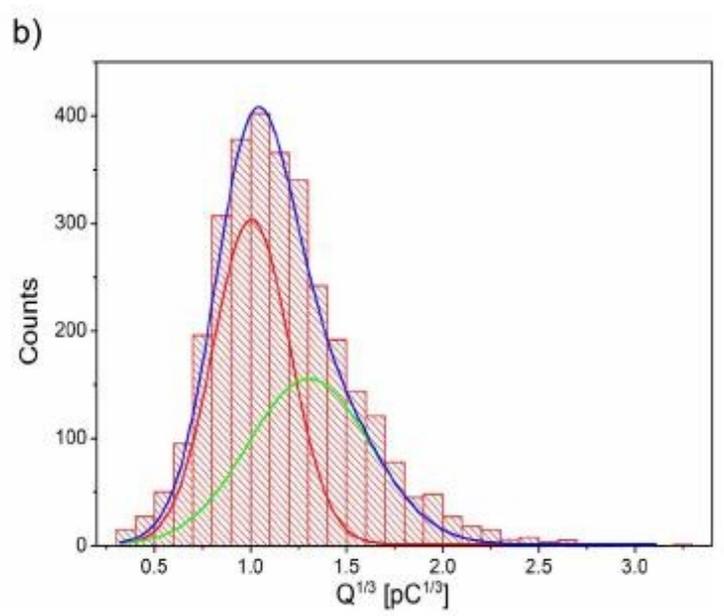